\documentclass[%
 preprint,
 12pt,
 amsmath,amssymb,
 aps,
]{revtex4-2}
\usepackage{siunitx}
\usepackage{graphicx}
\usepackage{dcolumn}
\usepackage{bm}
\usepackage{color}

\begin{document}


\title{Qubit-based clock synchronization for QKD systems using a Bayesian approach}

\author{Roderick D. Cochran}
\email{cochran.467@osu.edu}
\author{Daniel J. Gauthier}
\affiliation{%
 Department of Physics, The Ohio State University, Columbus, OH, 43210 USA
}%




\date{\today}

\begin{abstract}
    Quantum key distribution (QKD) systems provide a method for two users to exchange a provably secure key.  Synchronizing the users' clocks is an essential step before a secure key can be distilled.  Qubit-based synchronization protocols directly use the transmitted quantum states to achieve synchronization and thus avoid the need for additional classical synchronization hardware.  Previous qubit-based synchronization protocols sacrifice secure key either directly or indirectly, and all known qubit-based synchronization protocols do not efficiently use all publicly available information published by the users.  Here, we introduce a Bayesian probabilistic algorithm that incorporates all published information to efficiently find the clock offset without sacrificing any secure key.  Additionally, the output of the algorithm is a probability, which allows us to quantify our confidence in the synchronization.  For demonstration purposes, we present a model system with accompanying simulations of an efficient three-state BB84 prepare-and-measure protocol with decoy states.  We use our algorithm to exploit the correlations between Alice’s published basis and mean photon number choices and Bob’s measurement outcomes to probabilistically determine the most likely clock offset.  We find that we can achieve a 95 percent synchronization confidence in only 4,140 communication bin widths, meaning we can tolerate clock drift approaching 1 part in 4,140 in this example when simulating this system with a dark count probability per communication bin width of $8 \times 10^{-4}$ and a received mean photon number of 0.01.
\end{abstract}

\maketitle



\section{\label{sec:level1} Introduction}
Introduced in 1984 \cite{bb84}, quantum key distribution (QKD) is a symmetric encryption protocol that promises unconditional information security founded on the fundamental laws of physics, rather than on the difficulty of computational problems.  Bennett and Brassard established the first QKD protocol (BB84), which used the polarization degree of freedom of single photons to transmit information.  Subsequently developed protocols have extended QKD to different types of systems \cite{ekert1991} and relaxed the requirement for a true single-photon source \cite{hklo05}, paving the way for practical implementations of quantum cryptography.

For the sake of concreteness, we consider a polarization-based prepare-and-measure protocol.  Here, one user (Alice) prepares and transmits a periodic sequence of quantum states with period $\tau_A$ encoded in at least two mutually unbiased orthonormal bases.  In our example system, we use two bases: horizontal/vertical (H/V) polarization and left circular/right (L/R) circular polarization.  We also use the decoy-state protocol where Alice occasionally sends the vacuum quantum state.  A second user (Bob), measures each quantum state randomly in one of the two bases and records the result.  After this measurement phase is complete, Alice and Bob publish their basis choices for each measurement and keep only the measurements where Bob registers a click with his single-photon counting detectors and they use the same basis.  This process, called sifting, allows distilling a raw key, which, after error correction and privacy amplification \cite{bennett1991}, becomes the secret classical key securely shared between Alice and Bob.  Our example system uses a pulsed stochastic photonic source with decoy states \cite{hklo05}, where the decoys are photonic wavepackets with a lower mean photon number.  To simplify the example system and make it more efficient, we only transmit one state in the monitoring basis, which gives an equivalent secure key rate in comparison to transmitting both states in this basis \cite{islam18,tamaki2014}.

A practical issue in quantum communication protocols is synchronizing Alice and Bob's two data streams.  If Bob does not know precisely when Alice begins data transmission, he must begin recording measurements early or else risk missing some of Alice’s transmission.  In either case, because signals do not arrive at Bob due to channel loss, and extraneous events are caused by stray light and detector dark counts, the first event Bob records is unlikely to be the first event Alice sends, resulting in some timing offset that must be determined.  Correcting this offset is an essential precursor to sifting: If Alice and Bob do not agree on the timing of the events, they will compare basis choices from different events, resulting in a high quantum bit error rate (QBER) and likely share no information.  In addition, determining which time bins correspond to Alice's wavepacket arrival and which do not allows timing-based noise filtering.

Further complicating the communication protocol is that the relative clock offset may not be a constant due to drift in the relative phase and frequency between the transmitter and receiver clocks.  Alice has a communication protocol temporal bin width $\tau_A$ that may be different from Bob's bin width $\tau_B$.  The timing offset between their clocks $\Delta$ at the $n^{th}$ communication time bin since the most recent clock synchronization is given by 
\begin{equation}
    \Delta = t_0 + (\tau_A-\tau_B)n + \varepsilon  \label{eq:Delta}
\end{equation}
for an initial timing offset $t_0$ and higher-order timing error $\varepsilon$.  In this way, small differences in clock frequencies can gradually change the clock offset so that a previously calculated synchronization is no longer valid.  Other timing errors, such as clock jitter and frequency drift, also contribute to the need for a more robust synchronization solution.  We denote the time over which synchronization is maintained as $T_b$ (mnemonic \textit{batch}), \textit{i.e.}, the time over which the error in $\Delta \ll \tau_A$ 

Clock synchronization is sometimes achieved by directly sending Alice’s clock signal to Bob over a separate channel via an optical link or using a radio-frequency signal \cite{dauria2020, korzh2015, yliu2010, pliu2019, walenta2014, dynes2016, sasaki2011, wang2014, vallone2015, bourgoin2015}.  However, this introduces additional hardware requirements and increases the cost and complexity of the setup.  One way to avoid these additional resource requirements is to use the quantum channel itself to transmit the information necessary to perform the synchronization \cite{calderaro2020,agnesi2020,avesani2021,Ho_2009}.  One such qubit-based synchronization protocol was introduced and demonstrated by Calderaro \textit{et al.} \cite{calderaro2020}.  Their protocol uses a dedicated clock-synchronization phase followed by a key distribution phase.  In the synchronization phase, a pre-agreed synchronization string is transmitted to Bob and the clocks are aligned during post-processing.  

Their procedure does not correct for clock frequency drift: It only addresses the initial session time offset and frequency difference.  If the clock frequencies are not consistent, this method only temporarily aligns until the clock drift becomes of-the-order-of the communication protocol temporal bin width $\tau_A$.  Correcting for clock frequency drift requires repeated synchronization/key distribution phases with a regularity that depends on the stability of the clocks used in the experiment.  This reduces the overall secure key rate because no QKD states can be sent while the synchronization states are being sent, which may result in zero key rate due to finite-key effects \cite{hayashi2007,lim2014}. 

Our method avoids these limitations by synchronizing the clocks using only information that is already publicly sent over the insecure classical channel by Alice and Bob for sifting and security analysis: The basis choices and the mean photon number of the transmitted signal.  Because we are transmitting only one state in the monitoring basis, the basis choices provide information about which of Bob’s measurement outcomes are more likely.  The decoy state choices, which determine Alice's mean photon number for each wavepacket, also contain information about Bob’s measurement outcomes. For example, Bob is unlikely to record any detections if Alice sends the vacuum decoy state.

By comparing this information to his measurement outcomes, Bob can probabilistically determine the timing offset.  To account for potential clock drift, Bob can perform this synchronization in batches of length $T_b$. Thus, Bob can find the up-to-date timing offset and ensure that the basis choices he publishes are properly lined up with the ones sent to him by Alice, but this requires an efficient analysis method to reduce the data requirements.  Of course, our approach as well as Calderaro's requires low enough channel loss so that there are enough events received by Bob over a drift interval as discussed below.  

Another example of a qubit-based synchronization protocol for a continuously-pumped entanglement-based QKD systems was introduced by Ho \textit{et al.} \cite{Ho_2009}.  Here, they correlate Alice and Bob's detection events without considering basis information.  Their synchronization method relies on Alice's knowledge that some communication time bins are empty (assuming essentially unit detection efficiency for Alice's setup) and hence Bob's corresponding time bin should also be empty.  There is a single dominant peak in the correlation function that identifies $\Delta$ assuming a large enough number of Bob's detection events.  Because the detection timing information must already be shared publicly, this strategy does not sacrifice any secure key.  This method fails when the probability of Alice generating a photon per communication time bin approaches unity because every time bin is likely to be filled and hence the correlation function will have multiple high-value peaks that creates timing ambiguity.  

In the next section, we outline our synchronization algorithm and its advantages, and derive a formula for the synchronization probability using Bayesian analysis.  In Sec.~3 introduce a model system, and in Sec.~4 we simulate data in this model system to demonstrate the effectiveness of our method.  In Sec.~5 we present our conclusions and the potential applicability of this work to other QKD systems.  

\section{Qubit-Based Synchronization Algorithm} \label{Sec:SynchAlgorithm}

Similar to previous approaches, our algorithm uses a cross-correlation of Alice's periodically transmitted data and Bob's received data to find the number of each type of event pairing, where the cross-correlation is computed efficiently using a Fast Fourier Transform (FFT).  One complication of a prepare-and-measure scheme is that Alice attempts to send a quantum state every communication time bin, corresponding to the high-photon-probability limit of the Ho \textit{et al.} \cite{Ho_2009} method discussed above.  This problem is addressed here using the decoy-state protocol \cite{hklo05}, which must be used anyway to prevent a photon-number-splitting attack.

Decoy states are sent by Alice randomly and correspond to wavepackets with a mean photon number smaller than the signal state and often includes sending the vacuum state.  The vacuum state is particularly effective in the synchronization process because Alice has high certainty that she sent no photons, limited by her ability to completely block the source.  Bob should then also see no photons, limited by the source of detection clicks from non-ideal effects such as detector dark counts, detector afterpulsing, stray light, and the bleed through of light from Alice's source.

Beyond the decoy states, there are additional sources of correlation that can be exploited to help improve the synchronization process.  For example, Alice's use of the efficient three-state protocol, where she only sends one state in the monitoring basis, gives useful information if Alice and Bob also share basis-state information, which is already required for sifting.  We use a Bayesian statistical method, described below, that uses all prior knowledge of the system characteristics, such as the state fidelities, the mean photon numbers, the channel loss, the fractional sorting of Bob’s device for the two bases, and the detector efficiency, we generate a lookup table of Bob’s detection probabilities for Alice’s different inputs.  With these, we can easily compute the synchronization probabilities of different possible offsets using Bayesian statistics.  Alice and Bob's data is most correlated when they are synchronized.

A significant advantage of our approach is that it does not sacrifice any secure key: We only use the information that is already sent publically over the insecure classical channel.  This is an improvement over synchronization protocols that share some fraction of the raw data for synchronization purposes, as well as protocols that have a dedicated clock-synchronization phase \cite{calderaro2020} during which no QKD states can be sent.  

Bayesian analysis is a logical choice for synthesizing all available information and using it to make accurate predictions about $\Delta$.  It also has the advantage that it predicts the probability that $\hat{\Delta}$ is the best estimate of synchronization offset.  This allows us to quantitatively express our level of confidence in the synchronization estimate.  Furthermore, the additional information we incorporate in the protocol allows us to make a decision with fewer received qubits, which makes the system more robust to clock drift.

\subsection{Synchronization Probability}

Here we will use the strings of Alice and Bob's data.  A string of Bob's data consists of the results of each of his detectors at each sampling bin.  Typically, Bob's strings are very sparse because there are many sampling bins in which he registers no detections.  A string of Alice's data consists of her published information at each sampling bin.  If the communication time bin width is greater than the sampling time bin width, Alice will have multiple string entries for each state she sends, each corresponding to what she is sending at that part of her duty cycle.  Determining the synchronization probability consists of comparing different strings of Bob’s data (starting at different temporal offsets) to strings of Alice's data and calculating which of Bob’s strings $D$ is most likely to be the one generated by Alice’s corresponding string.  We determine, for a particular string of Bob’s, the probability that it could have been generated by Alice’s published string.

Mathematically, we phrase this as the likelihood $p(D|S)$ of generating Bob’s string $D$ given the assumption that its generating string is the one Alice has published, denoted by $S$.  The uninformed assumption, which we will denote as $\bar{S}$, is that Bob’s string $D$ has been generated by a random string other than Alice’s published string (from some other portion of Alice's sent data), with the stipulation that the other string is also periodic.  This mathematical framework will consider a subset of Alice's data of $N$ sampling bin widths compared against a subset of Bob's data of $N+M$ sampling bin widths, meaning there will be $M$ possible offsets to consider.

To begin in our protocol formalism, we note that $D$ is a string of length $M+N$ of Bob's measurements at each sampling bin (including sampling bins where no detections were received). Each measurement $B_i$ in Bob's string consists of the click or no-click results at all of Bob's detectors.  Bob's string $D$ can be written as
\begin{equation}
D = \{B_{1}, ... , B_{M+N}\},
\end{equation}
which we can rewrite as
\begin{equation}
D = \{B_{1}, D'\},
\end{equation}
where 
\begin{equation}
D' = \{B_{2}, ... , B_{M+N}\}.
\end{equation}
We prefer to write the likelihood $p(D|S)$ in terms of known quantities such as the $p(B_1|S)$, the conditional probability of a time bin measurement $B_1$ given $S$.  Using this notation, $p(D|S)$ is given by
\begin{equation}
p(D|S) = p(B_{1},D'|S) = p(B_{1}|D',S)p(D'|S),
\end{equation}
where the final equality is a result of the product rule.  Because we have assumed that $B_{1}$ is generated from Alice’s string, knowing $D'$ gives us no additional information about $B_1$.  At best, it informs us whether $S$ true, which is already assumed; the bits are otherwise independent because Alice’s sequence is random.  Using these observations, we obtain  
\begin{equation}
p(B_{1}|D',S) = p(B_{1}|S),
\end{equation}
and, by extension, 
\begin{equation} \label{eq:lik_prod}
p(D|S) = \prod\limits_{i=1}^{N+M} p(B_{i}|S),
\end{equation}
allowing us to write the likelihood as the product of the measurement probabilities at each sampling bin.
We note that even in the example where Alice only sends one state in the monitoring basis, Bob must still measure both states in each basis to detect potential eavesdropper attacks \cite{tamaki2014,islam18}.  For computational ease, we also determine each sampling bin measurement probability as the product of the probabilities of the outcomes at the four different detectors $b_{\ell}$, which are given by
\begin{equation}
p(B_i|S) = \prod\limits_{\ell=1}^{4} p(b_{\ell}|S)
\end{equation}
Again, because the detector events are assumed to be generated by independent random processes, these probabilities can be considered independent when the generating string is known.

When the generating string is not known (under the uninformed assumption $\bar{S}$), the detection probabilities can be approximated as independent when the received mean photon number is low.  Because the synchronization task is most difficult in low-signal regimes, we use this approximation going forward.  Thus,
\begin{equation}
p(D|\bar{S}) = \prod\limits_{i=1}^{N+M} p(B_{i}|\bar{S}) \label{eq:nlik_prod}
\end{equation}
and
\begin{equation}
p(B_i|\bar{S}) = \prod\limits_{\ell=1}^{4} p(b_{\ell}|\bar{S}).
\end{equation}
For a given input from Alice, each of Bob’s four detectors has an opportunity to detect a photon above the detection clicks arising from non-ideal behaviors.  Naturally, we will use our knowledge of the system (the state fidelities, the quality of the polarization sorting, the dark count rates, the detector efficiencies, and the signal and decoy received mean photon number) to estimate the detection probabilities as accurately and efficiently as possible.  Using a lookup table of the detection probabilities for the different inputs from Alice, these likelihoods can be calculated using standard statistical methods.

However, the likelihood of generating $D$ from Alice’s published string is not the same as the probability that Alice’s published string is the one that generated $D$, which is given by $p(S|D)$ and is the most relevant quantity to determine synchronization.  Bayes’ theorem allows us to rewrite this quantity, called the posterior, as
\begin{equation}
p(S|D) = \frac{p(D|S)p(S)}{p(D)}. \label{bayes}
\end{equation}
In addition, we must also include the information that we expect exactly one correct synchronization offset (not just one on average).

To formulate the problem as an exclusive synchronization, we must find the probability that some discreet timing offset, given by the time-bin index $j$, is the correct synchronization offset, and that all the other offsets are incorrect.  In other words, the probability that, for a given string of length $N$ published by Alice, all the measurements before the $j^{th}$ bin are generated randomly, the measurements from $j$ to $j+N$ are generated from Alice's published string, and the measurements after $j+N$ are generated randomly.  Under these assumptions, we can write $p(B_1,...,B_{M+N}|S_j)$ as a product of the likelihoods of these three sections as
\begin{equation} \label{eq:3part}
    p(B_1,...,B_{M+N}|S_j) = p(B_1,...,B_{j-1}|\bar{S_j})p(B_j,...,B_{j+N}|S_j)p(B_{j+N+1},...,B_{M+N}|\bar{S_j}).
\end{equation}
Here we introduce $\bar{S_j}$, the assumption that the data is produced by a random string other than the synchronization string in question, but one with the same phase (\textit{i.e.}, the signal arrives at the same time bin in each period as it does for $S_j$).

We can find the conditional probability for matching Alice's string to Bob's string at a potential synchronization index $j$ in this framework using Eq.~\ref{bayes}, which gives
\begin{equation}
p(S_j|B_1,...,B_{M+N}) = \dfrac{p(B_1,...,B_{M+N}|S_j)p(S_j)}{p(B_1,...,B_{M+N})}. 
\label{eq:likelihoodPSD}
\end{equation}
Equation~\ref{eq:likelihoodPSD} is our main result and is the quantity of interest to identify clock synchronization between Alice and Bob.  We determine the optimum synchronization index based on the value of $j$ that maximizes this quantity, and the quantity itself gives us our confidence in that choice.

The denominator in Eq.~\ref{eq:likelihoodPSD} can be written in terms of known quantities using marginalization.  Marginalization consists of rewriting a probability as a sum of the comprehensive conditional probabilities; in this case, the different possible synchronization indices written as
\begin{equation}
p(S_j|B_1,...,B_{M+N}) = \dfrac{p(B_1,...,B_{M+N}|S_j)p(S_j)}{\displaystyle\sum_{i=1}^{M} p(B_1,...,B_{M+N}|S_{i}) p(S_{i})},
\end{equation}
where the $i$ denotes the other potential synchronization indices.

To evaluate Eq.~\ref{eq:likelihoodPSD}, we the likelihoods 
$p(B_1,...,B_{M+N}|S_j)$ and $p(B_1,...,B_{M+N}|S_i)$ can be determined using Eqs.~\ref{eq:lik_prod}, \ref{eq:nlik_prod} and, \ref{eq:3part}.  The quantity $p(S_j)$, called the prior, is the $\textit{ad hoc}$ probability that $D$ corresponds to Alice’s published string.  That is, that we are at the correct synchronization index. We use a uniform prior, which assumes each candidate has a na\"ive $1/M$ probability of being the correct one given that we have $M$ candidate indices, which means that
\begin{equation}
p(S_{i}) = p(S_{j}) = \dfrac{1}{M}
\end{equation}
so that the prior terms cancel, giving us
\begin{equation}
p(S_j|B_1,...,B_{M+N}) = \dfrac{p(B_1,...,B_{M+N}|S_j)}{\displaystyle\sum_{i=1}^{M} p(B_1,...,B_{M+N}|S_{i})}.
\end{equation}
Next, we apply Eq.~\ref{eq:3part} to obtain
\begin{eqnarray}
 p(S_j|B_1,...,B_{M+N}) = \\ \nonumber
 &&\hspace*{-1.5cm} \dfrac{p(B_1,...,B_{j-1}|\bar{S_j})p(B_j,...,B_{j+N}|S_j)p(B_{j+N+1},...,B_{M+N}|\bar{S_j})}{\displaystyle\sum_{i=1}^{M} p(B_1,...,B_{i-1}|\bar{S_i})p(B_i,...,B_{i+N}|S_i)p(B_{i+N+1},...,B_{M+N}|\bar{S_i})}
\end{eqnarray}
and use Eqs.~\ref{eq:lik_prod} and \ref{eq:nlik_prod} (of which the latter uses a low received mean photon number approximation) to write everything in terms of known quantities as
\begin{equation}\label{eq:master}
 p(S_j|B_1,...,B_{M+N}) \approx \dfrac{\displaystyle\prod_{k=1}^{j-1}p(B_k|\bar{S_j})\displaystyle\prod_{k=j}^{j+N}p(B_k|S_j)\displaystyle\prod_{k=j+N+1}^{M+N}p(B_k|\bar{S_j})}{\displaystyle\sum_{i=1}^{M}\Bigg( \displaystyle\prod_{k=1}^{i-1}p(B_k|\bar{S_i})\displaystyle\prod_{k=i}^{i+N}p(B_k|S_i)\displaystyle\prod_{k=i+N+1}^{M+N}p(B_k|\bar{S_i})\Bigg)}
\end{equation}

Equation~\ref{eq:master} is our master equation for the synchronization probability of an index $j$.  The numerator consists of the probability of an $N$-length string of Bob's data starting at $j$ being produced by Alice's published string, along with the probability that the remaining data was produced by an unknown string of Alice's data.  The denominator sums this same quantity over all possible synchronization indices, ensuring normalization.  We take the value of $j$ that maximizes this quantity to be the optimum synchronization index, and the value of $p(S_j|B_1,...,B_{M+N})$ gives us the probability that we are correct.  We can compute this conditional probability using FFTs to count the number of each unique bin measurement along with a lookup table of the probabilities of the events.

\section{Model System}

\begin{figure}[h!]
\includegraphics[width=9cm]{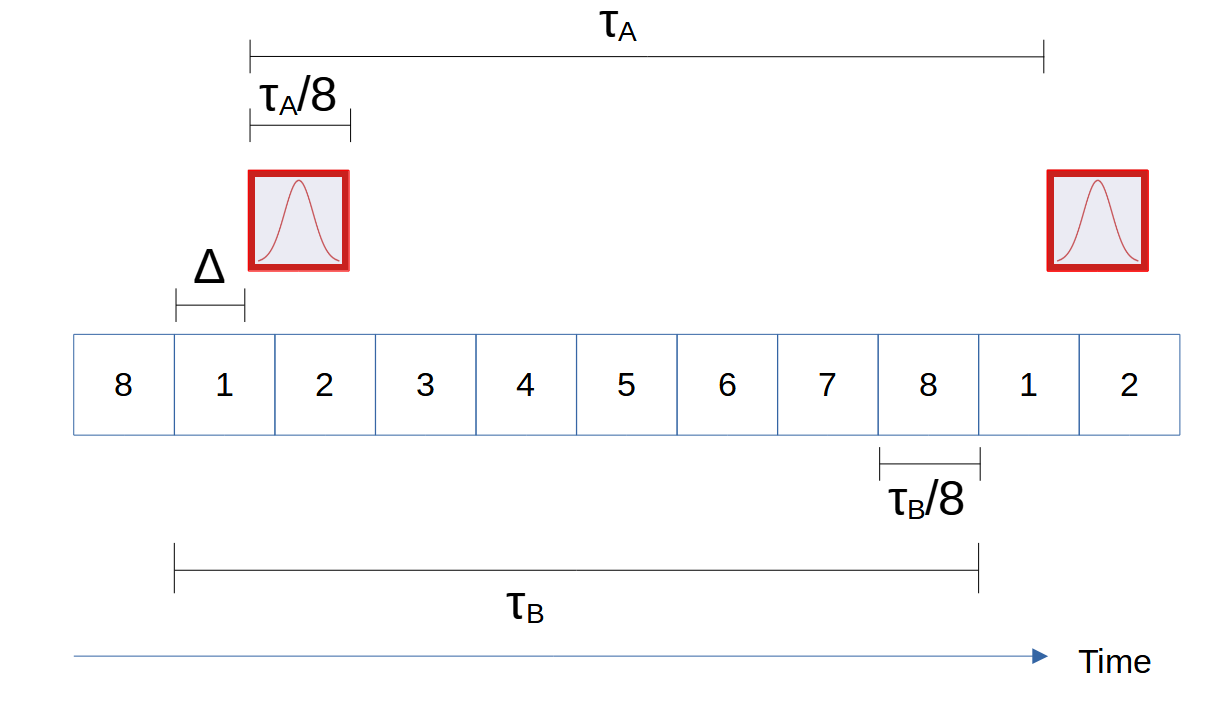}
\caption{Illustration of the relative times used in the QKD protocol. Here, the signal (red) straddles bins 1-2 due to an offset of $\Delta$, and we do not consider bins 3-8.  We take $\tau_A = \tau_B$, which is approximately correct for a short enough data subset.
}\label{Fig:timing}
\end{figure}

To illustrate our protocol, we simulate a model QKD system using a polarization-based prepare-and-measure protocol with decoy states and only sending one state in the monitoring basis.  We set Alice's repetition rate to be $f_A=1/\tau_A$ and a wavepacket duration of $\Delta t = \tau_A/m$ with $m=8$ for a duty cycle of 12.5 percent.  We set Bob's sampling rate to $n f_A$ with $n=8$ so that his sample period is matched to the wavepacket duration. These conditions are illustrated in Fig.~\ref{Fig:timing}. Alice generates a pseudorandom sequence such that four quantum states L/R/H and a vacuum decoy state (a decoy state with mean photon number equal to zero) are sent in equal parts on average. 

For our numerical experiments, we simulate a QKD session by generating data that emulates the state preparation and measurement, including aspects such as the \textit{received} mean photon number $\mu$, probability of a detector dark count $d$ over one communication bin width $\tau_A$, and variation in $\Delta$ due to clock drift, assumed to be constant over $T_b$.  This allowed us to test how these factors impact the synchronization performance.  We assume a transmitted mean photon number $\mu_A = 1$ where the received mean photon number $\mu = \eta \mu_A$ for a channel transmission $\eta$.  While this $\mu_A$ is on the upper end of values used in typical experiments, it allows us to explore the performance and limitations of our algorithm at or beyond the greatest received mean photon number one would realistically use: $\mu_A  = 1$ with zero loss.

Assuming a Poisson distribution for Alice's source, the probability of Bob registering a click $p(click)$ over a period $\tau_A$ at a particular detector $\ell$ is given by
\begin{equation}\label{eq:pclick}
    p(click, \ell) = 1 - (1-d) e^{\mu_{\ell}}
\end{equation}
where $\mu_{\ell}$ is the mean photon number received by detector $\ell$.  The portion of the total mean photon number $\mu$ that goes to the different detectors depends on which polarization state is sent.  We use ideal BB84 sorting in our model system so that all states have an equal chance of being measured in either basis.  States measured in the same basis as they are prepared are detected accurately, while states measured in the opposite basis have an equal chance of being measured in either opposite-basis state.  For example, if Alice prepared an H-state that Bob receives $\mu=0.8$, Bob's measures $\mu_H=0.4$, $\mu_v=0$, and $\mu_L=\mu_R=0.2$.-
In the low-$\mu$ limit, Eq.~\ref{eq:pclick} can be approximated as

We assume that the observation window is long enough so that the $p$'s and $\mu$'s can be estimated accurately from the finite number of observations.  This means the average click probability can be extracted from the Bob's raw data and we rewrite Eq.~\ref{eq:pclick} as a function of ${p(click, \ell)}$ so that
\begin{equation}
    {\mu_{\ell}} = ln \bigg(\dfrac{1-d}{1-{p(click, \ell)}}\bigg)
\end{equation}
and we estimate the received mean photon number of a signal state as 
\begin{equation}\label{eq:pclickbar}
    {\mu} = \dfrac{4}{3} \displaystyle \sum_{\ell=1}^4 ln \bigg(\dfrac{1-d}{1-{p(click, \ell)}}\bigg),
\end{equation}
which just sums the average mean photon number of the different detectors and multiplies by a factor of 4/3 to account for the fact that we are sending vacuum states 25\% of the time.

We divide the data set into subsets duration $T_b$ and perform synchronization and sifting on each subset.  Bob can record up to 8 events (each of which may or may not include a detection event or dark count) assuming that the detector deadtime is less than Bob's sampling time.  However, because the clocks can only be synchronized to a resolution of Bob's sampling bin width, we expect Alice’s wavepacket to straddle 2 bins as illustrated in Fig.~\ref{Fig:timing}, with the end bins only having a partial wavepacket.  The remaining 6 bins only contain dark counts, which can be discarded after we determine $\Delta$ to reduce noise.  This amounts to detector time-gating in the post-analysis. 

We assume that Bob begins recording before Alice begins transmitting, and continues to record after she stops sending, so our received data is bookended by low signal regions.  We find a best-fit step function to identify where the transmission begins and ends, which gives us a coarse approximation of the synchronization index.  For a range of different string lengths $N$ that determine the number of sampling bin widths in each synchronization subset, we examine a window of $M = 4,000$ nearby potential synchronization indices.  This value is chosen based on the typical precision of the coarse approximation of the synchronization given by the best-fit step function.

\section{Synchronization Simulations}

\begin{figure}[h!]
\includegraphics[width=9.5cm]{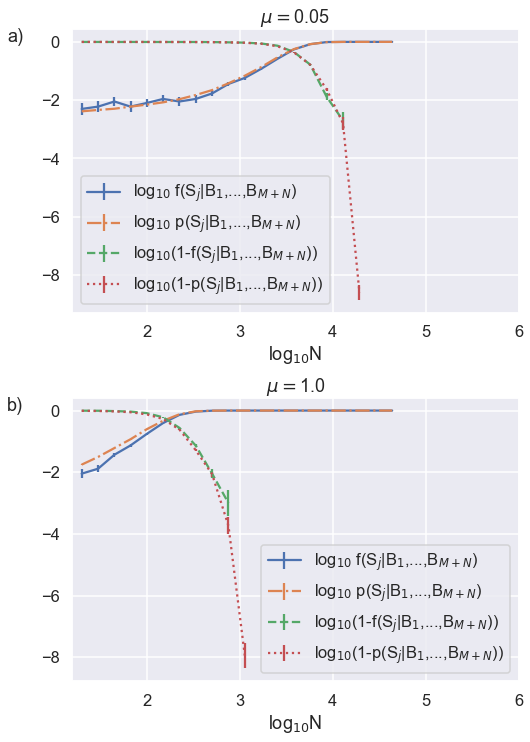}
\caption{\label{fig:prob_test} Bob's required data record length needed to determine synchronization for two different channel transmissions of (a) $\eta = 0.05$, corresponding to $\mu=0.05$ and (b) $\eta = 1$, corresponding to $\mu=1$.  We also show the probability of not obtaining synchronization, which better highlights transition to high-certainty synchronization. 
}
\end{figure}

To verify that our algorithm returns an accurate probability of synchronization, we run 1,000 simulated trials with a known synchronization index and compared the average calculated probability of synchronization $p(S_j|B_1,...,B_{M+N})$ to the average rate of finding the correct index, which we denote by $f(S_j|B_1,...,B_{M+N})$, in Fig.~\ref{fig:prob_test}.  If our model is accurate, then $p(S_j|B_1,...,B_{M+N}) \sim f(S_j|B_1,...,B_{M+N})$, in which case we can take $p(S_j|B_1,...,B_{M+N})$ to be a reliable metric for quantifying our confidence in obtaining the correct $\Delta$.

We see that $p(S_j|B_1,...,B_{M+N}) \sim f(S_j|B_1,...,B_{M+N})$ to within our errorbars for moderate channel loss (Fig.~\ref{fig:prob_test}a).  However, $p(S_j|B_1,...,B_{M+N})$ is consistently larger than $f(S_j|B_1,...,B_{M+N})$ for the case of zero channel loss (Fig.~\ref{fig:prob_test}b), a condition that is unlikely to be encountered in an experiment but highlights the limitation of our algorithm.  This result is not surprising given that our derivation given in Sec.~\ref{Sec:SynchAlgorithm} assumes low $\mu$ to arrive at Eq.~\ref{eq:nlik_prod}.  Assuming a transmitted mean photon number of 1, Fig.~\ref{fig:prob_test} (b) corresponds to a zero channel loss system.  This represents an upper limit on $\mu$ encountered in a typical decoy state protocol where $\mu_A \lesssim 1$ and thus, also serves as a lower bound on the accuracy of our calculated synchronization probability.

\begin{figure}[h!]
\includegraphics[width=11cm]{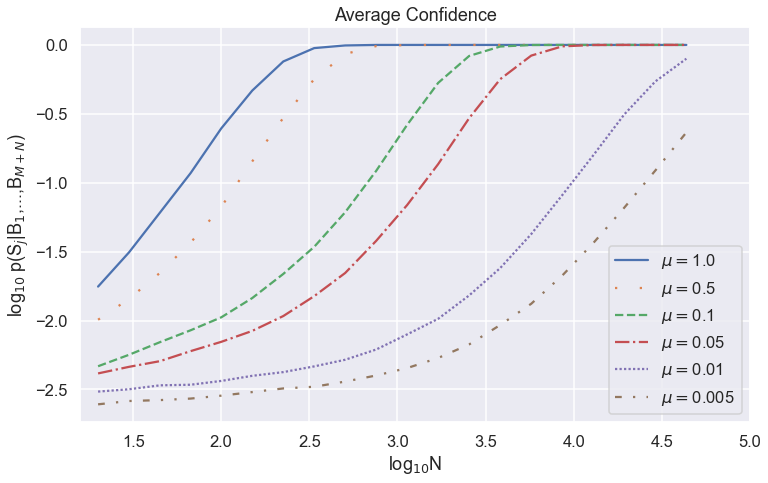}
\caption{\label{fig:sync_param} Average calculated synchronization probability as a function of string length on a logarithmic scale for different received mean photon numbers.  The probability of registering a dark count during one communication bin width is $d = 8\times10^{-4}$.
}
\end{figure}

A lower received mean photon number means a lower density of detected events.  Because detected events provide more information than no-detection events, a lower $\mu$ requires us to consider a larger set of sampling bin widths $N$ to achieve the same synchronization confidence.  Despite the fact that $p(S_j|B_1,...,B_{M+N})$ does not match $f(S_j|B_1,...,B_{M+N})$ as well at higher values of $\mu$, we can still achieve equivalent average values of $f(S_j|B_1,...,B_{M+N})$ at lower values of $N$.  This fact is also illustrated in Fig.~\ref{fig:sync_param}, where we see a direct correlation between $\mu$ and $N$ at which the synchronization probabilities converge to one: The higher values of $\mu$ converge at lower values of $N$.  

\begin{figure}[h!]
\includegraphics[width=8.5cm]{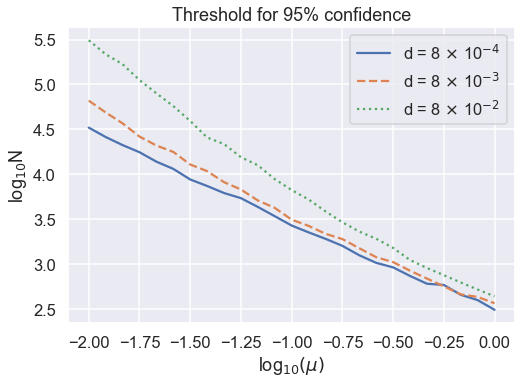}
\caption{\label{fig:sync_95} Dependence of string length threshold to achieve 95 percent synchronization confidence on received mean photon number on a logarithmic scale, parameterized by different dark count probabilities.
}
\end{figure}

Another way to view this relation between $\mu$, $N$, and $p(S_j|B_1,...,B_{M+N})$ is to consider the string length $N$ required to achieve a particular synchronization confidence as a function of $\mu$ as shown in Fig.~\ref{fig:sync_95}.  For high $\mu$ and low $N$, we observe an approximately linear relation between $log_{10}\mu$ and $log_{10}N$ with a slope of $\sim$ -1, which means that $N~1/\mu$.  For lower $\mu$, where there are fewer events and dark counts play a larger role, the probability curves exhibit steeper slopes, demonstrating that synchronization becomes increasingly difficult.  This data can be used to estimate whether it is possible to synchronize over an experimentally measured temporal block length $T_b$ and, if it is possible, how low a value of $\mu$ can be tolerated while still synchronizing reliably.  As a concrete example, Bob needs 33,110 sampling bin widths, or about 4,140 communication bin widths, to achieve a 95\% confidence for clock synchronization for $\mu=0.01$ and $d=8 \times 10^{-4}$. This means we can tolerate clock drifts approaching 1 part in 4,140 because our method assumes that the clock drift is much less than one communication bin width.

\section{Conclusions}
In conclusion, we develop a novel probabilistic approach to qubit-based clock synchronization using Bayesian analysis.  By exploiting correlations between information Alice shares publicly, such as basis and decoy state choices, and Bob's detection events, we can find the correct synchronization clock offset without sacrificing any secret key.  Additionally, our algorithm is more robust to noise, loss, and clock drift in comparison to other protocols by incorporating all publicly available information using the Bayesian framework.  Finally, we demonstrate that our algorithm is successful and robust using a simulated BB84 communication scheme, which confirms that our synchronization metric corresponds to the probability of synchronization, especially in the low-$\mu$ limit.  Our algorithm is applicable to other QKD systems that use other degrees-of-freedom of the photon for which it is possible to divulge some timing information. 

\section*{Acknowledgements}

This material is based on research sponsored by NASA under grant 80NSSC20K0629 and the Air Force Research Laboratory and the Southwestern Council for Higher Education under agreement FA8650-19-2-9300.  The U.S. Government is authorized to reproduce and distribute reprints for Governmental purposes notwithstanding any copyright notation thereon.  The views and conclusions contained herein are those of the authors and should not be interpreted as necessarily representing the official policies or endorsements, either expressed or implied, of NASA, the Southwestern Council for Higher Education and the Air Force Research Laboratory (AFRL), or the U.S. Government.  R.D.C. acknowledge discussions of the Bayesian analysis with Richard Furnstahl.  All code and data used in simulations is publicly available on GitHub:\\ \url{https://github.com/roderickdcochran/qubit_based_synchronization}



\bibliography{sync_refs}

\end{document}